# Cold Dark Matter Resuscitated?

Martin White, Douglas Scott, Joseph Silk, and Marc Davis
*Center for Particle Astrophysics and Departments of Astronomy and Physics*
*University of California, Berkeley CA 94720-7304*

May 1995

**ABSTRACT**
The Cold Dark Matter (CDM) model has an elegant simplicity which makes it very predictive, but when its parameters are fixed at their 'canonical' values its predictions are in conflict with observational data. There is, however, much leeway in the initial conditions within the CDM framework. We advocate a re-examination of the CDM model, taking into account modest variation of parameters from their canonical values. We find that CDM models with $n = 0.8$–$0.9$ and $h = 0.45$–$0.50$ can fit the available data. Our "best fit" CDM model has $n = 0.9$, $h = 0.45$ and $C_2^T/C_2^S = 0.7$. We discuss the current state of observations which could definitely rule out this model.

*'Bring out your dead. Bring out your dead.'*
*'I'm not dead yet!'*
   *– Monty Python*

## 1 INTRODUCTION

The most successful of all cosmological models for large-scale structure has surely been that of inflationary ($\Omega_{\rm tot} = 1$, initially scale-invariant fluctuations) cold dark matter, otherwise known as CDM (Peebles 1982; Blumenthal et al. 1984; Davis et al. 1985). It has been thoroughly examined via N-body and SPH simulations on scales ranging from galaxies to clusters and superclusters. Only in the post-*COBE* (Smoot et al. 1992) era, when the normalization of the fluctuations in the linear regime has finally been determined, has the theory become vulnerable to attack on several fronts. The overwhelming opinion among cosmologists currently holds that CDM is now in flagrant conflict with observational data, notably on scales of $\sim 1 - 10$ Mpc (e.g. (Davis et al. 1992; Liddle & Lyth 1993; Ostriker 1993)). With the popular alternatives of either a cosmological constant-dominated or an open universe requiring initial conditions that are (for many physicists) distasteful to contemplate, we think it behooves us to carefully reconsider the CDM model before discarding it.

Indeed, we will show that remaining within the framework of the standard CDM model but with judicious tuning of the canonical parameters, cold dark matter is worthy of serious reconsideration. New observations will be needed, described below, before a definitive resolution of the astronomical issues can be accomplished. We note in passing that confirmation of a small ($\sim$ few eV) neutrino mass would provide another means of maintaining a viable cold dark matter-dominated universe, in this case with a $\sim 20\%$ admixture of hot dark matter. However in the absence of an experimental measurement of neutrino mass, we regard the coincidence required in order of magnitude between $\Omega_\nu$ and $\Omega_{\rm CDM}$ to be unappealing. In any case we feel that a mixed dark matter model is best considered as a minor variation on the theme of a CDM universe.

We will consider models which keep the attractive features of standard CDM, while incorporating only mild deviations from the standard parameters. The main attraction of CDM is its simplicity: (1) $\Omega_{\rm tot} = 1$, with the dominant mass in the form of massive particles whose interactions and velocity dispersion are negligible; (2) adiabatic fluctuations drawn from a Gaussian random distribution; and (3) initial conditions which can arise naturally from inflation. We believe that there are models which satisfy these requirements while being in reasonable agreement with the data.

## 2 COSMOLOGICAL DATA

Given the enormous breadth of available cosmological data, which pieces of information does the CDM model need to address? Making such a choice is inevitably a subjective process that involves taking a somewhat skeptical view of many published claims, in part because of uncertain systematics such as those encountered in recent Hubble constant determinations (cf. (Freedman et al. 1994; Nugent et al. 1995)). We will also avoid the issue of whether there is any evidence for or against having $\Omega_0 = 1$ on the largest scales as measured by velocity flows (Dekel 1994; Strauss & Willick 1995; Shaya, Peebles & Tully 1995); if it becomes clear that $\Omega_0 < 1$ then the models we discuss will no longer be viable. Instead of these classical cosmological parameters, we will use data which measure the power over some range of scales: $\sigma(r)$ is the rms overdensity within spheres of size $r$, and the dimensionless power spectrum is $\Delta^2(k) \equiv d\sigma_\infty^2/d\ln k = k^3 P(k)/(2\pi^2)$.

We will specifically concentrate on the following list of constraints:

(i) the *COBE* normalization at the largest scales;
(ii) a measurement of the shape of the spectrum at large scales;
(iii) determination of the normalization of the fluctuations on cluster scales;
(iv) the potential energy as measured by velocities on Mpc scales; and
(v) an indication that there may be an acoustic peak in the CMB fluctuations.

To commence, we remark that the theoretically cleanest number is the *COBE* (Bennet et al. 1994) normalization of the potential fluctuations on the horizon scale (Bunn, Scott & White 1995; Stompor, Gorski & Banday 1995). We impose the *COBE* constraint by normalizing all of our models directly to the *COBE* 2-year data using the method of (White & Bunn 1995). While the *COBE* data gives information on the shape of the power spectrum at large scales, the change in likelihood across the range of parameters we consider is not significant, so we do not include it.

Secondly we want to use information from large scales about the break in the spectrum. Frequently this is specified in terms of a shape parameter, $\Gamma \simeq \Omega_0 h$, but we prefer to use a measure more closely tied to the data. Since information on scales $\gtrsim 100\,h^{-1}{\rm Mpc}$ is sparse, most current $\Gamma$ limits are really statements about relative amounts of "large" to small scale power. We choose to consider the ratio between the contribution to the rms power at large ($\sim 30\,h^{-1}{\rm Mpc}$) scales, $\Delta_{0.05} \equiv \Delta(k = 0.05\,h{\rm Mpc}^{-1})$, to that at small scales: $\sigma_8$ (the rms overdensity at $8\,h^{-1}{\rm Mpc}$, which probes power at wavenumber $\simeq 0.2 h\,{\rm Mpc}^{-1}$)[*] both in *linear* theory. A value for $\Delta_{0.05}/\sigma_8$ can be found from the data compilation of (Peacock & Dodds 1994): $\Delta_{0.05}/\sigma_8 = 0.27 \pm 0.02$. This value for $\Delta_{0.05}/\sigma_8$ is consistent with the value determined from power spectral analysis of a range of surveys once corrections are made for redshift space distortions. For standard CDM this ratio is 0.22, which is the usual statement that this model has insufficient large-scale power. If we fix $n=1$ and use the fitting function for $T(k)$ of (Efstathiou 1990), we find $\Delta_{0.05}/\sigma_8 \simeq 0.21 + 0.22(1/2 - \Gamma)$ for $\Gamma$ near $1/2$. While there is certainly more information about large-scale structure than this, in general a smooth $P(k)$ that passes this test, and others outlined below, provides a reasonable fit to most of the data (see Figure 2).

We next need to confront the overall amplitude of the *mass* fluctuations on cluster scales: $\sigma_8$. The best way to estimate this is probably by means of cluster abundances, which can be related to the amplitude of the fluctuations in linear theory through the Press-Schechter ansatz. Recent determinations (Bond & Myers 1991; White, Efstathiou &

Frenk 1993; Carlberg et al. 1994) lead to $\sigma_8 \simeq 0.5 - 0.8$, where the principle uncertainty is in the assumed cluster mass ($\sigma_8 \sim m^{0.4}$ in the Press-Schechter theory), though the formalism employed and data used give rise to some of the spread. A lower limit on $\sigma_8$ can be obtained by considering rare (high mass or temperature) clusters at redshift $z > 0$, though this is not a constraint for the models considered here. Since for optical galaxies $(\sigma_{8,g})^2 \simeq 0.9 \pm 0.05$ (Loveday et al. 1992; Loveday et al. 1995), this constraint can also be phrased in terms of bias: $b \equiv \sigma_{8,g}/\sigma_8 \simeq 1.2$–$2$.

One area where the CDM model has run into trouble has been its prediction of velocities on Mpc scales which are too "hot" (Suto, Cen & Ostriker 1992; Gelb & Bertschinger 1994; Schlegel et al. 1994; Nolthenius, Klypin & Primack 1994). This is a problem for most $\Omega_0 = 1$ models, which predict hot velocity fields whereas the small scale velocity field appears locally to be quite cold (Peebles 1992). Rather than focus on the relative pair dispersion, $\sigma_{12}$, which is dominated by clusters, we shall consider the small scale rms velocity of individual galaxies: $v_{\rm rms}$. Redshift catalogues typically give 3-dimensional measurements of $v_{\rm rms} \sim 350\,{\rm km\,s^{-1}}$ (Miller & Davis 1995), though refined analyses of N-body simulations and large redshift catalogues will be required for more precise constraints. To address this we calculate the 2nd moment of the mass correlation function, i.e. $J_2(1\,h^{-1}{\rm Mpc})$, in non-linear theory using the fitting function for $P(k)$ of (Mo, Jain & White 1995). This gives an estimate of the potentials in a filtered version of the Layzer-Irvine or "cosmic energy" equation, which can be related to the three-dimensional rms velocity on Mpc scales through (Peebles 1980; Davis & Peebles 1983)

$$v_{\rm rms}^2 \simeq \frac{6\Omega_0}{7 + n_{\rm eff}}\,J_2 H_0^2. \qquad (1)$$

Note that since conventionally $J_2(1\,h^{-1}{\rm Mpc})$ is measured in $(h^{-1}{\rm Mpc})^2$, the dependence on $H_0^2$ cancels in $v_{\rm rms}$. On purely dimensional grounds $J_2(1\,h^{-1}{\rm Mpc})$ should also be related to the pairwise velocity dispersion squared, $\sigma_{12}^2$, determined by the cosmic virial theorem, if scaling solutions are used to relate the three-point function to the two-point function (as is often done). We note the accurate calculation of the pairwise velocity dispersion is not unambiguous (e.g. (Zurek et al. 1994; Marzke et al. 1995)). Given the above, it seems that a reduction in $J_2$ on Mpc scales by a factor $> 3$ (relative to standard CDM) is desirable. We find that this it is possible to reduce $J_2$ by a factor 3–4, but it is unclear whether this is enough.

At essentially the same scale, but constraining the power spectrum from the other direction, are abundances of quasars and damped Ly$\alpha$ systems at high redshift. We can regard the data as leading to a lower limit on the rms mass fluctuation at $1\,h^{-1}{\rm Mpc}$ in linear theory: $\sigma_1$. This can be used to compute abundances based on Press-Schechter (or peak-patch) estimates and to obtain a feel for the redshift at which structures would form: $(1 + z_{\rm form}) \approx \sigma_1$ (but see (Katz et al. 1994) for simulations of early structure formation). Ideally we would like to keep $\sigma_1 \gtrsim 3$, though we shall see that early structure formation is not a problem for any of our models.

Finally, there is the observationally immature but theoretically crucial issue of the acoustic peak in degree-scale CMB fluctuations (Scott, Silk & White 1995). Although the possibility of systematic errors and foreground contamination remain an issue, the data at face value require the ratio of the large angle to degree-scale fluctuations $D_{200}/D_{10} \gtrsim 2$ (with a rather large error bar) where $D_\ell \simeq \ell^2 C_\ell$. This last constraint will be very important if strengthened, since the peak goes away if we tilt too much or introduce too much tensor contribution. However, this depends to a large extent on $\Omega_B$: the peak becomes higher as $\Omega_B$ increases. If new measurements of the primordial deuterium abundance give a value of a few $\times 10^{-4}$ (Songaila et al 1994), forcing $\Omega_B$ to the lower end of the nucleosynthesis range, then the presence of a Doppler peak would put a much stronger constraint on the allowed $n$ and hence $\sigma_8$.

---

[*] This may be compared with the "excess power" defined in (Wright et al. 1992), which probes similar scales but is harder to obtain from the published data.

Should a variant of CDM be able to fit these numbers, it will simultaneously fit many of the other less rigorous tests summarized elsewhere (e.g. (Silk 1994; Bond 1994; Bahcall 1994)). We regard the strongest of these additional tests as being the requirement that the Hubble constant should not be much lower than $H_0 \approx 50 \text{ km s}^{-1} \text{ Mpc}^{-1}$, and that the age of the universe should be $\gtrsim 12$ Gyr. It appears to be possible to match all of the above constraints by considering small deviations from the parameters of standard CDM.

## 3 THE CDM MODEL

In almost every 'realistic' inflationary model, the spectral index, $n$, of the primordial fluctuation spectrum is only *approximately* unity. Indeed (unless there is tuning of parameters e.g. (Barrow & Liddle 1993)) the limit of *exact* scale invariance corresponds to an exactly de Sitter inflationary phase (which would never end) and is unphysical. This is apparent in the divergence of the amplitude of the adiabatic fluctuations generated by inflation as $n \to 1$. Two classes of inflationary models generate fluctuation spectra which are almost exact power laws: natural inflation (Adams et al. 1993) and power law or extended inflation (Lucchin & Matarrese 1985; La & Steinhardt 1989; Steinhardt & Accetta 1990; Kolb, Salopek & Turner 1990; Lyth & Stewart 1992). The other classes generally have spectra which are only approximately power laws, with the corrections becoming more important as the spectrum departs from scale invariance (see e.g. (Stewart & Lyth 1994)). For simplicity, we will focus on power law initial fluctuation spectra, though we emphasize that a scale-dependent spectral index is not unlikely if the fluctuations are generated by inflation (Steinhardt & Turner 1984; Kofman & Linde 1987; Kofman & Pogosyan 1988; Salopek, Bond & Bardeen 1989; Liddle & Turner 1994). Many authors have stressed that having $n$ somewhat below unity is a desirable feature (e.g. (Bond 1994; Scott, Silk & White 1995)), and certainly tilted CDM models have been considered before (e.g. (Cen et al. 1995; Liddle et al. 1992; Muciaccia et al. 1993)). However such studies have usually been in terms of relatively extreme tilt, i.e. $n \lesssim 0.7$, and without using the accurate normalization on large scales provided by the *COBE* 2-year data.

Power law inflation models predict a component of tensor fluctuations whose amplitude is related to that of the scalar component by an amount proportional to $(1-n)$ (Liddle & Lyth 1992; Davis et al. 1992; Stewart & Lyth 1994), while natural inflation models predict a negligibly small tensor component (Adams et al. 1993). The tensor fluctuations affect only the predictions of the model for large scale CMB anisotropies, but this affects the normalization of the model in terms of *COBE*. The ratio of tensor and scalar components is usually expressed in terms of their contribution to the quadrupole CMB anisotropy. For power law inflation $C_2^T/C_2^S = 7(1-n)$ while for natural inflation $C_2^T/C_2^S \approx 0$. We examine tilted CDM models both with and without tensor contributions and restrict the tensor contribution to that predicted by power law inflation.

We attempt to fit a wide range of observational data by allowing $n$, $C_2^T/C_2^S$ and $h$ to deviate by only small amounts from their canonical values. Specifically, we assume that $\Omega_B h^2 = 0.02$ from BBN (Copi et al. 1995; Krauss & Kernan 1995), and allow $h$ to range between 0.45 and 0.55 and $n$ to range between 0.8 and 1. Higher values of $n$ and $h$ could clearly be considered, but these give worse and worse fits to the data. Adopting a lower value of $\Omega_B$ would make it more difficult to keep $D_{200}/D_{10} \gtrsim 2$, since we are generally decreasing $n$ to get $\sigma_8$ to observationally acceptable levels. Though in principle it would also become harder to fit the high redshift damped Ly$\alpha$ abundance with lower values of $\Omega_B$, our models produce enough small scale power that this is not a serious constraint.

Power spectra for our models were calculated numerically with a Boltzmann code. The *transfer functions*, including the effects of our assumed baryon content, can be fit by the form of (Efstathiou 1990) with $\Gamma = 0.39, 0.45, 0.51$, for $h = 0.45, 0.50, 0.55$ respectively[†]. Fitting functions for various parameters discussed above, over the range $0.45 \leq h \leq 0.55$ and $0.8 \leq n \leq 1$, are given in Table 1. The error on $\sigma_8$ and $\sigma_1$ simply from the *COBE* normalization is 7.5%, while it is 15% for $J_2(1\,h^{-1}\text{Mpc})$. The expressions in Table 1 are formally accurate to $\lesssim 5\%$, but for $J_2$ have additional uncertainty due to the non-linear nature of the calculation.

## 4 DISCUSSION

Figure 1 shows the contours of $\Delta_{0.05}/\sigma_8$, $\sigma_8$, $\sigma_1$ and $J_2(1\,h^{-1}\text{Mpc})$ vs $n$ and $h$ for models without and with tensors, respectively. To have $\sigma_8 \leq 0.8$ we are forced into the bottom left region of the plots. Choosing round numbers our best fit models have either: (1) $h \simeq 0.45$, $n \simeq 0.8$, $C_2^T/C_2^S \simeq 0$; or (2) $h \simeq 0.45$, $n \simeq 0.9$, $C_2^T/C_2^S \simeq 0.7$. The values from Figure 1 are listed in Table 2 for these two models and for "standard" CDM. Also shown in Figure 2 are the matter power spectra for these two models plus an MDM model (with $\Omega_\nu = 0.2$) and a $\Lambda$CDM model (with $\Omega_0 = 1 - \Omega_\Lambda = 0.3$, $h = 0.8$, $n = 0.95$ and $C_2^T/C_2^S = 7(1-n)$ (Scott, Silk & White 1995)) for comparison. The data points are an estimate of the *linear* $P(k)$ from (Peacock & Dodds 1994), and are plotted assuming $\Omega_0 = 1$. [The non-linear power spectrum we have used to estimate $J_2(1\,h^{-1}\text{Mpc})$ and $v_{\text{rms}}$ is also shown.] There is an uncertainty of at least 20% in the overall normalization of these data points. In addition the amplitude of the points scales as $\Omega_0^{-0.3}$, so we have multiplied the $\Lambda$CDM curve by $0.3^{0.3}$ to compensate. The radiation power spectra, $D_\ell \equiv \ell(\ell+1)C_\ell$ vs $\ell$, for our two models are shown in Figure 3.

---

[†] Note: for $n < 1$ the "effective" $\Gamma$ for $P(k)$ would be lower than for $T(k)$.

Our models are able to accomodate the measure of $\Delta_{0.05}/\sigma_8$ from (Peacock & Dodds 1994). Not surprisingly they are also in accord with other measures of "shape" such as the zero-crossing of the cluster correlation function ($\xi(r_{cc}) = 0$ for $r_{cc} \sim 50\, h^{-1}$ Mpc, (Klypin & Rhee 1994)) and are reasonable fits to the APM angular correlation function $w(\theta)$. For $h \leq 0.5$ we can reduce $\sigma_8$ to acceptable levels, though as can be seen in Figure 1 we find it difficult to obtain $\sigma_8 \leq 0.7$ without the inclusion of tensors.

The amplitude of the mass fluctuations on $1\, h^{-1}$ Mpc scales is consistent with estimates from the abundances of quasar and damped Ly$\alpha$ systems at high redshift, very roughly $\sigma_1 \gtrsim 3$ today (Haehnelt 1993; Ma & Bertschinger 1994; Kauffmann & Charlot 1994; Mo & Miralda-Escudé 1994; Subramanian & Padmanabhan 1994; Liddle & Lyth 1995). These are rather soft numbers to aim at, but the models we consider have enough small scale power to form objects early, in contrast with models like MDM where small scale power in the hot component is exponentially damped.

Figure 1 or Table 2 shows that a factor of 3–4 reduction in $J_2(1\, h^{-1}\, \mathrm{Mpc})$ over the standard CDM number is possible, though it is difficult to obtain more suppression than this and still get high $\sigma_1$. (The values of $J_3(1\, h^{-1}\, \mathrm{Mpc})$ show a similar dependence on the input parameters, so if $J_2$ is in the acceptable range, we expect $J_3$ will be also.) The "hot" velocity fields on Mpc scales are potentially the biggest problem for these models, in common with most other $\Omega_0 = 1$ models. In order to address the question of possible velocity bias (Carlberg & Couchman 1989), and to study this problem in more detail, we are investigating the velocity fields for these models in cosmological N-body simulations.

## 5 CONCLUSIONS

We propose that the two models in Table 2 are currently the 'best-buy' versions of CDM, and contend that these models, in which the parameters are varied by $\sim 10\%$ from their canonical values, should be the standard by which CDM is judged. Before abandoning the appealing simplicity of standard CDM in favour of $\Lambda$ or massive neutrinos (or more extreme modifications, such as isocurvature fluctuations or topological defects), it is these inflationary-inspired CDM models which should be confronted with observational data.

Despite our belief that these mild variants of CDM are a reasonable fit to all currently available data, there are certainly future tests which could definitively rule them out. The age determination is the one classical test that is relevant because it is largely the recent flurry of activity in Hubble constant determinations that is forcing reconsideration of low-$\Omega$ universes: if measurements of $h$ become firm at values like 0.7–0.8, then there is no way to save CDM without at least introducing a cosmological constant. However, conflicting claims abound, and we are presently reluctant to place any more emphasis on $h = 0.8 \pm 0.08$ than $h = 0.5 \pm 0.05$.

The present situation on large scale flows is ambiguous (Dekel 1994; Strauss & Willick 1995; Shaya, Peebles & Tully 1995), but improved data should yield a definitive measure of $\beta \equiv \Omega_0^{0.6}\sigma_8$. Given a reliable determination of $\sigma_8$, the degeneracy between $\Omega_0$ and $\sigma_8$ could be broken. If indications that $\Omega_0 < 1$ from these considerations are ultimately confirmed, then we will have to give up the aesthetic appeal of an Einstein-de Sitter universe. Another strong test will be the so-called 'baryon catastrophe' in clusters (White et al. 1993): if the fraction of baryons in clusters implies that $\Omega_B/\Omega_0 \simeq 0.2$ then this is difficult for all high $\Omega_0$ models. Here one currently runs into uncertainty in the determination of cluster masses as measured by the X-ray emitting gas. Lensing studies of clusters will be an important check on possible systematic errors in the available mass estimates.

From the point of view of the microwave background, the most damaging observations would be: if the acoustic peak became a firm detection at smaller angular scales, since it is not possible to achieve this with a flat geometry; or if the peak has height $\gg 4$, which is hard to achieve with $n < 1$ and no cosmological constant. Another clear test which may perhaps be performed in the next few years is a more definitive measurement of the matter power spectrum shape. If the break in the power spectrum is really at larger scales than predicted by CDM, mild variations of the CDM hypothesis no longer suffice, and a more drastic revision of our current best-fit model for large-scale structure is required.

## ACKNOWLEDGEMENTS

This work was supported in part by grants from the NSF and the DOE.

| Quantity | 1 | $\tilde{h}$ | $\tilde{n}$ | $\tilde{h}\tilde{n}$ | $\tilde{n}^2$ | $\tilde{h}\tilde{n}^2$ |
|---|---|---|---|---|---|---|
| | | | Coefficient of | | | |
| $\Delta_{0.05}/\sigma_8$ | 0.22 | -0.28 | 0.17 | -0.16 | | |
| $\sigma_8$ | 1.33 | 3.31 | -2.40 | -6.27 | — | — |
| $\sigma_1$ | 11.1 | 31.6 | -34.5 | -79.9 | 41.4 | — |
| $J_2(1\,h^{-1}\,{\rm Mpc})$ | 205 | 1060 | -738 | -2899 | 536 | — |
| $\sigma_8$ | 1.33 | 3.19 | -5.54 | -9.25 | 9.58 | — |
| $\sigma_1$ | 11.1 | 30.3 | -55.0 | -100. | 103. | — |
| $J_2(1\,h^{-1}\,{\rm Mpc})$ | 203 | 1034 | -1493 | -5753 | 3355 | 8971 |

**Table 1.** Fitting formulae for measures of power as a function of $\tilde{h} \equiv h - 1/2$ and $\tilde{n} \equiv 1 - n$, valid to 5% over the range $0.45 \le h \le 0.55$ and $0.8 \le n \le 1$. The error on the normalization from *COBE* is 7.5% for $\sigma_8$ and $\sigma_1$ and 15% for $J_2(1\,h^{-1}\,{\rm Mpc})$. The fits are divided into two cases: without (upper) and with (lower) tensors. The shape ratio is independent of the tensor contribution.

| Quantity | "Obs" | $C_2^T = 0$ | $C_2^T > 0$ | "CDM" |
|---|---|---|---|---|
| $V_{60}$ | $327 \pm 82$ | 282 | 254 | 356 |
| $\Delta_{0.05}/\sigma_8$ | $0.27 \pm 0.02$ | 0.27 | 0.25 | 0.22 |
| $\sigma_8$ | 0.5–0.8 | 0.75 | 0.77 | 1.34 |
| $\sigma_1$ | $> 3$–4 | 5.0 | 5.6 | 11 |
| $v_{\rm rms}$ | $< 600$ | 740 | 775 | 1400 |
| $D_{200}/D_{10}$ | 2–10 | 4.0 | 3.3 | 5.5 |

**Table 2.** Power spectrum measures for our preferred CDM models: $h = 0.45$, $n = 0.8$ (without tensors) and $h = 0.45$, $n = 0.9$ (with tensors). The values of $v_{\rm rms}$ (km s$^{-1}$) come from $J_2$ and Eq. (1), assuming $n_{\rm eff} = -1$. The "Observational" numbers come from papers cited in the text.

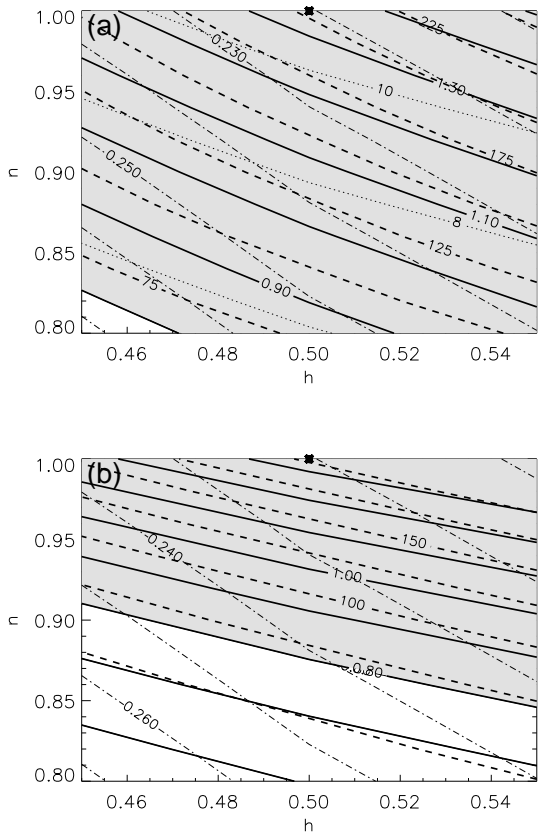

**Figure 1.** Contours of $\sigma_8$ (solid), $\Delta(0.05)/\sigma_8$ (dot-dashed), $\sigma_1$ (dotted) and $J_2(1\,h^{-1}\,{\rm Mpc})$ (dashed) for models with (panel b) and without (panel a) tensors. The contour of $\sigma_1$ has been suppressed in panel (b) for ease of viewing. The shaded region shows $\sigma_8 \geq 0.8$ and the "standard" Cold Dark Matter model is marked by an asterisk. Our best fit model has (a) $h \simeq 0.45$, $n \simeq 0.8$, $C_2^T/C_2^S = 0$ and (b) $h \simeq 0.45$, $n \simeq 0.9$, $C_2^T/C_2^S = 0.7$.

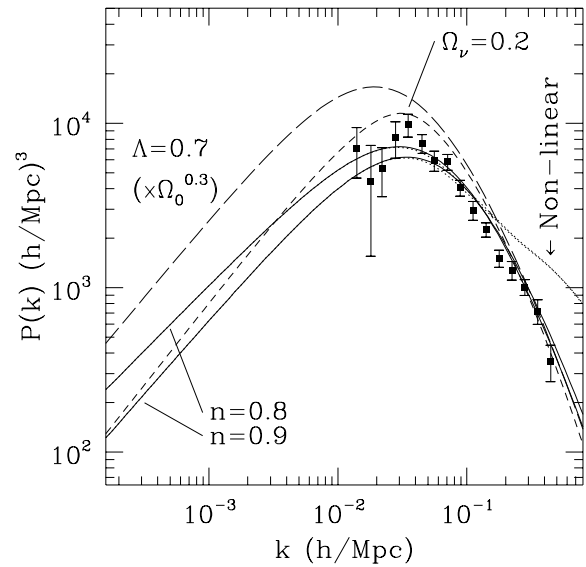

**Figure 2.** The matter power spectra for our CDM models (solid), along with an MDM model (short dashed) and $\Lambda$CDM model (long dashed) for comparison. The $\Lambda$CDM model has been multiplied by $\Omega_0^{0.3}$ to allow comparison with the data points (from Peacock & Dodds (1994)). The non-linear $P(k)$ for the CDM models is shown dotted.

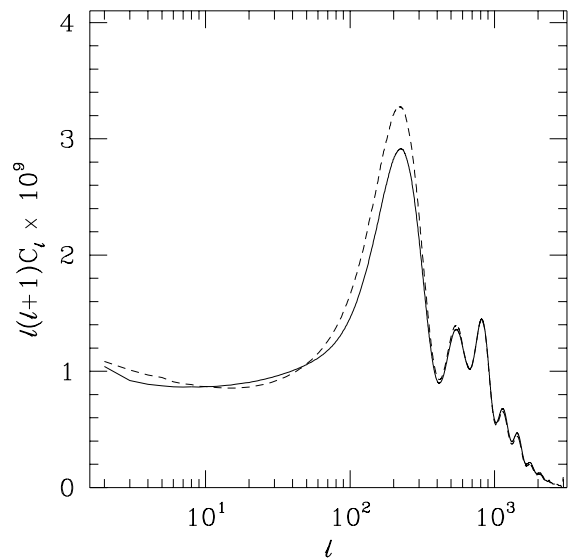

**Figure 3.** The radiation power spectra for our CDM models with (solid) and without (dashed) a tensor component, normalized to the 2-year *COBE* data.